\documentclass[%
 reprint,
showpacs,preprintnumbers,
amsmath,amssymb,
aps,
prl,
longbibliography
]{revtex4-1}

\usepackage{epsfig}
\usepackage{array}
\usepackage{multirow}
\usepackage{amsmath}
\usepackage{braket}
\usepackage{graphicx}
\usepackage{color}

\newcommand{\beginsupplement}{%
        \setcounter{table}{0}
        \renewcommand{\thetable}{S\arabic{table}}%
        \setcounter{figure}{0}
        \renewcommand{\thefigure}{S\arabic{figure}}%
     }

\begin{document}

\title{Breakdown of the Chiral Anomaly in Weyl Semimetals in a Strong Magnetic Field}

\author{Pilkwang Kim}
\author{Ji Hoon Ryoo}
\author{Cheol-Hwan Park}
\email{cheolhwan@snu.ac.kr}
\affiliation{Department of Physics, Seoul National University, Seoul 08826, Korea}

\date{\today}

\begin{abstract}

The low-energy quasiparticles of Weyl semimetals are a condensed-matter realization
of the Weyl fermions introduced in relativistic field
theory.
Chiral anomaly, the nonconservation of the chiral charge 
under parallel electric and magnetic 
fields, is
arguably the most important phenomenon of Weyl semimetals
and has been explained 
as an imbalance between the occupancies of the gapless,
zeroth Landau levels with opposite chiralities.
This widely accepted picture has served as the basis for subsequent
studies.
Here we report the breakdown of the chiral anomaly 
in Weyl semimetals in a strong magnetic field
based on {\it ab initio} calculations.
A sizable energy gap that depends sensitively on the direction 
of the magnetic field may open up
due to the mixing of the zeroth Landau levels associated with
the opposite-chirality Weyl points that
are away from each other in the Brillouin zone.
Our study
provides a theoretical framework for understanding a wide range of
phenomena closely related to the chiral anomaly in topological semimetals, 
such as magnetotransport, 
thermoelectric responses, 
and plasmons, to name a few.

\end{abstract}

\maketitle

Although Weyl fermions or chiral fermions 
are basic building blocks for relativistic theories, 
there are no known massless fermions with a definite chirality
among the elementary particles in nature
since neutrinos were found to be massive.
Very recently, however, it was found that some condensed-matter systems,
so-called Weyl semimetals, host Weyl fermions 
as low-energy quasiparticles~\cite{PhysRevB.83.205101}.  
Weyl semimetals have Weyl points in the Brillouin zone around which 
the energy versus momentum relation is linear.  
Shortly after theoretical
predictions in 2015~\cite{Huang:2015ic,PhysRevX.5.011029},
TaAs, NbAs, TaP, and NbP have been
confirmed to be Weyl semimetals by angle-resolved 
photoemission spectroscopy 
experiments~\cite{Xu613,PhysRevX.5.031013,
Xu:2015ky,Xue1501092,0256-307X-32-10-107101}.

  \begin{figure*}
  \centering
  \includegraphics[width=0.85\textwidth]{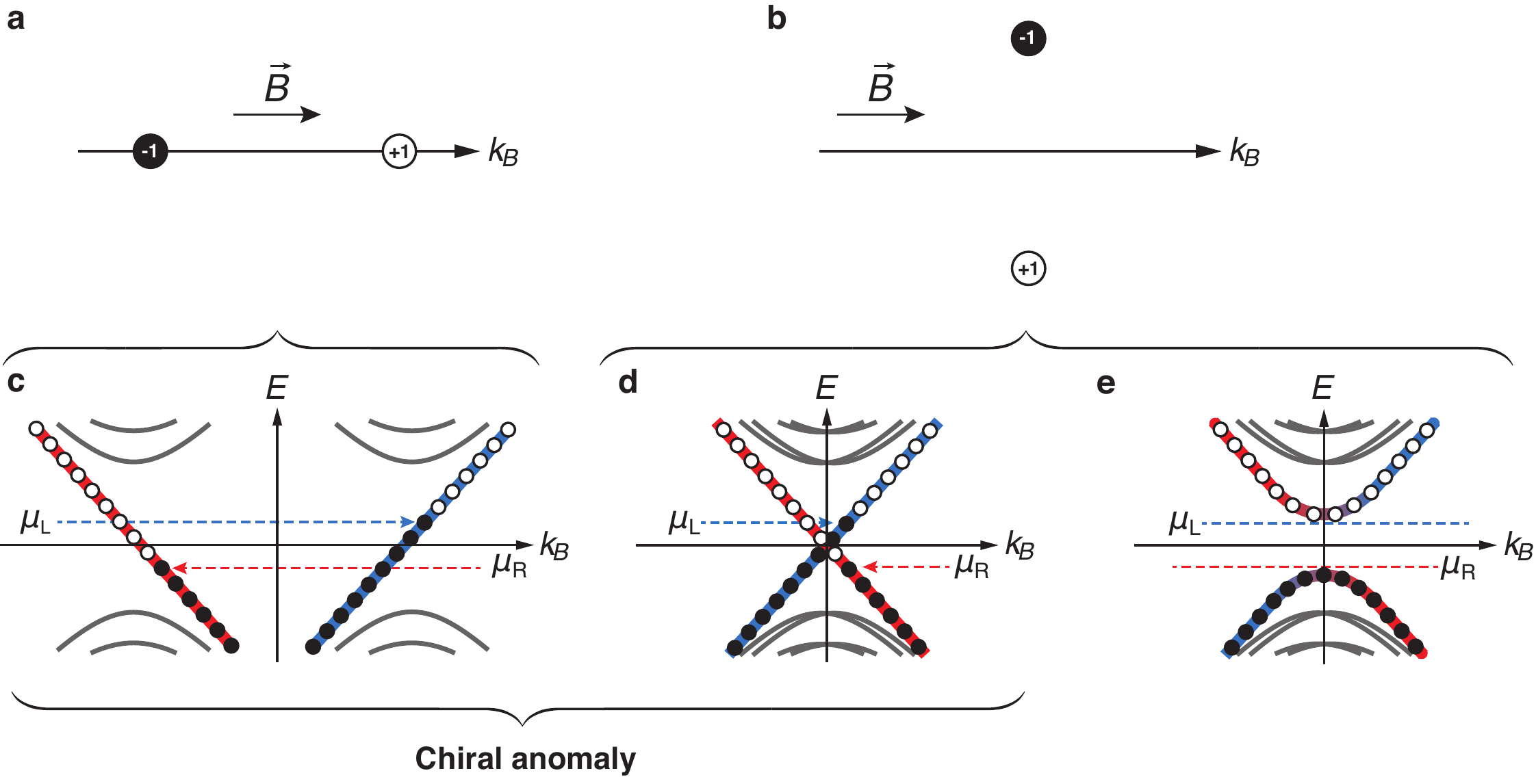}
  \caption{
  {\bf Chiral anomaly in Weyl semimetals and its breakdown.}  
  {\bf a} and {\bf b},
  Schematics of Weyl points in a magnetic field. 
  $k_{B}$ is the wavevector along the direction of the magnetic field. 
  Encircled numbers are the chiral charges of the Weyl points. 
  Two Weyl points with opposite chiralities lie along a line 
  parallel ({\bf a}) or perpendicular ({\bf b}) to the magnetic field.
  {\bf c}, A schematic of the Landau-level spectrum of Weyl fermions 
  with chirality $+1$ and $-1$ in the presence 
  of parallel electric and magnetic field.  
  Occupied and unoccupied states are shown 
  as black and white discs, respectively.   
  ${\mu_{\rm L}}$ and ${\mu_{\rm R}}$ are the chemical potentials 
  of the left and right electrodes and are associated with 
  the charge carriers moving to the right and to the left, respectively.  
  Here, the separation of the two Weyl points along $k_{B}$ is large 
  ({\bf a}).  
  Note that the populations of the Weyl fermions 
  with chirality $+1$ and $-1$ are different, which is chiral anomaly.  
  {\bf d}, The same schematic as in {\bf c} 
  but for a case with a small separation 
  in $k_{B}$ between the Weyl points ({\bf b}).  
  As in {\bf c}, the populations of the Weyl fermions 
  with chirality $+1$ and $-1$ are different.  
  {\bf e}, A similar schematic as in {\bf d} 
  but for a case with an energy gap.  
  Here the chiral anomaly breaks down.
  }
  \label{Fig1}
  \end{figure*}

The chiral anomaly, which is relevant 
to bulk transport~\cite{NIELSEN1983389},
is one of the two most intriguing phenomena
of Weyl semimetals, the other being the Fermi arcs appearing in the
surface electron spectrum~\cite{PhysRevB.83.205101}.
In relativistic field theory, 
chiral anomaly refers to the nonconservation 
of chiral charge in the presence of 
parallel electric and magnetic fields
and was first introduced to explain 
the two-photon decay of 
the neutral pion~\cite{PhysRev.177.2426,Bell:1969bi}.

In 1983, Nielsen and Ninomiya predicted that
the chiral anomaly in condensed-matter systems
manifests itself as negative 
longitudinal magnetoresistance~\cite{NIELSEN1983389}.
This prediction is based on 
the characteristic linear energy-vs-wavevector
dispersions of the chiral zeroth Landau levels (ZLLs).  
An electric field applied parallel to the magnetic field 
induces the charge imbalance between the ZLLs
with opposite chiralities (Figs.~1a and~1c). 
Although the chiral ZLLs may or may not intersect with each other
depending on how the two Weyl points 
align with the magnetic field (Figs.~1a and~1b),
the explanation by Nielsen and Ninomiya~\cite{NIELSEN1983389}
is valid in all cases (Figs.~1c and~1d).
All theoretical and experimental studies 
on the phenomena in Weyl semimetals related to the chiral anomaly
are based on this traditional picture~\cite{PhysRevB.85.241101, PhysRevB.86.115133,
PhysRevB.88.104412, PhysRevX.4.031035, 
PhysRevLett.113.247203, PhysRevB.91.245157, 
PhysRevX.5.031023, Shekhar:2015io, Zhang:2016cv, Arnold:2016hl,
PhysRevB.90.165115, Hirschberger:2016fb, 
PhysRevB.89.245103, PhysRevB.91.035114}.
Especially, the negative longitudinal magnetoresistance
that was originally suggested by Nielsen and Ninomiya~\cite{NIELSEN1983389}
has been observed in many experiments 
and interpreted as a signature of the chiral anomaly
in Weyl semimetals~\cite{PhysRevX.5.031023, Zhang:2016cv}.

  \begin{figure*}
  \centering
  \includegraphics[width=0.95\textwidth]{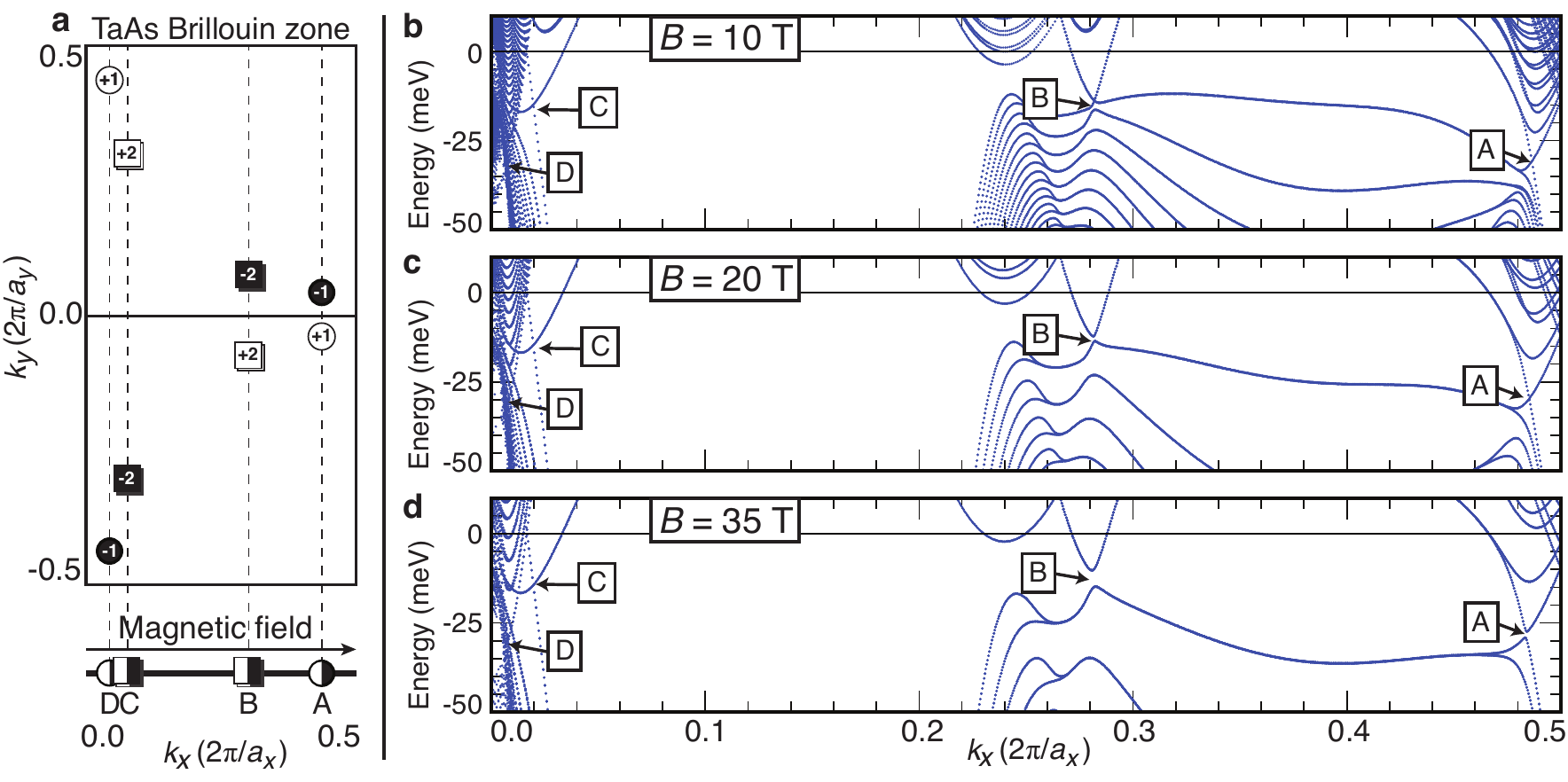}
  \caption{
  {\bf Landau levels of TaAs.}
  {\bf a}, A schematic (not to scale) of the projected Weyl points 
  and their chiral charges (the enclosed numbers).  
  Half of the Brillouin zone ($k_{x}>0$) is shown.  
  Discs and squares represent two different classes of Weyl points.  
  {\bf b}-{\bf d}, Landau-level spectra of TaAs 
  in a magnetic field applied along $x$ 
  obtained from {\it ab~initio} tight-binding calculations.  
  The zero of energy is set at the Fermi level before the application of the magnetic field.
  ${\rm A}$, ${\rm B}$, ${\rm C}$, and ${\rm D}$ denote the corresponding Weyl points shown in {\bf a}.  
  }
  \label{Fig2}
  \end{figure*}

In this paper, we report the breakdown of the chiral anomaly 
in Weyl semimetals in a strong magnetic field %
based on comprehensive {\it ab initio} calculations 
on the Landau levels of TaAs, NbAs, TaP, and NbP. 
In particular, we find that 
an energy gap may open up 
due to the mixing of the ZLLs associated with
the Weyl points with opposite chiralities 
that are separated in the Brillouin zone.
The size of the energy gap strongly varies
with the strength and direction 
of the applied magnetic field.

  \begin{figure}
  \includegraphics[width=1\columnwidth]
  {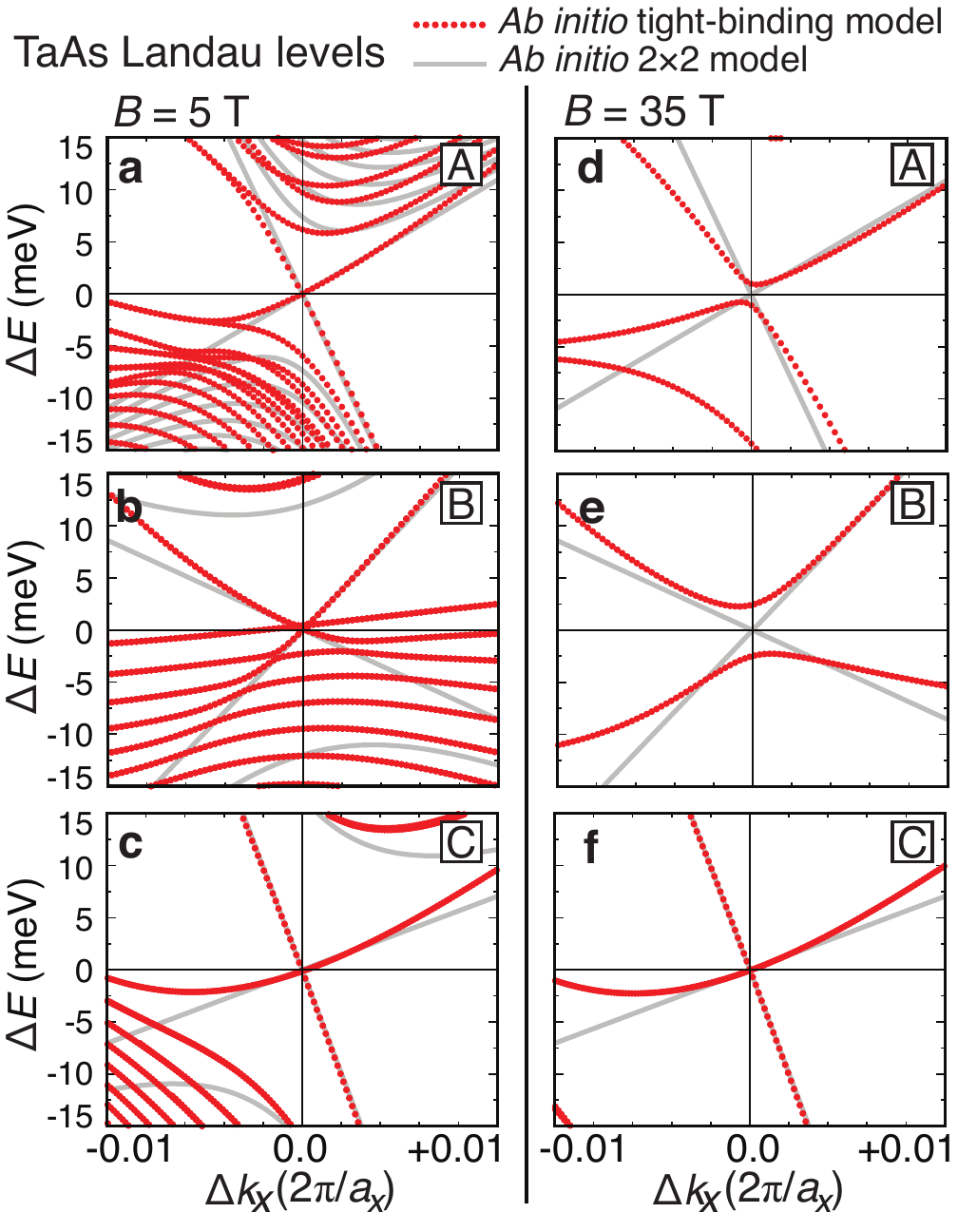}
  \caption{
  {\bf Magnetic-field-induced gap opening in TaAs.}
  Landau level spectra around ${\rm A}$ ({\bf a} and {\bf d}), 
  ${\rm B}$ ({\bf b} and {\bf e}), and ${\rm C}$ ({\bf c} and {\bf f}), 
  at $B=5$~T ({\bf a}-{\bf c}) and $B=35$~T ({\bf d}-{\bf f}). 
  Red discs and solid grey curves show the results 
  obtained from the {\it ab~initio} tight-binding model 
  and those obtained from 
  the {\it ab~initio} 2$\times$2 model, respectively.  
  The zero of energy is set at the mid-point of 
  the higher- and lower-energy zeroth Landau levels. 
  }
  \label{Fig3}
  \end{figure}

We first consider the Landau levels of TaAs 
in a magnetic field along $x$ 
(see Supplementary Figs.~1 and~2 for 
the crystal structure and the electronic band structure, respectively~\cite{Note_supp}).
The energy-vs-wavevector relation of Landau levels 
is dispersive only along 
the direction of the applied magnetic field.
There are four positive $k_{x}$ values at which 
the Landau-level states are associated with 
the Weyl points of TaAs (${\rm A}$, ${\rm B}$, 
${\rm C}$, and ${\rm D}$ in Fig.~2a). 
In a relatively weak magnetic field ($B \le 10$~T
 where $B$ is the strength of the magnetic field), 
the {\it seemingly} gapless ZLLs are 
clearly visible near ${\rm A}$, ${\rm B}$,  and
${\rm C}$
(Figs.~2b and~3a-3c).

Remarkably, energy gaps open up at 
the crossing of the ZLL energy bands 
around A and B as the magnetic field increases 
(Figs.~2c, 2d, 3d, and 3e).
The {\it gapped} ZLLs are simultaneously 
associated with the two Weyl points of opposite chiralities
that are separated in the Brillouin zone. 
The charge carriers occupying 
the gapped ZLLs are therefore not chiral.

We emphasize that this gap opening is totally different from
the previously reported gap opening 
in Dirac semimetals~\cite{PhysRevLett.115.176404,
Potter:2016jq,Liu:2016fw}.
It was found that an external magnetic field 
that breaks the rotational symmetry protecting the Dirac
semimetallic phase induces an energy gap in the ZLLs 
of a Dirac semimetal~\cite{PhysRevLett.115.176404,
Potter:2016jq,Liu:2016fw}.
However, according to those studies, the ZLLs of a Weyl
semimetal should remain gapless.

  \begin{figure}
  \includegraphics[width=1\columnwidth]
  {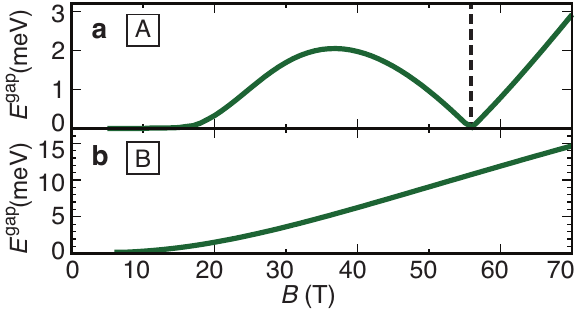}
  \caption{
  {\bf Band inversion and non-trivial dependence 
  of the energy gap 
  on the strength of the magnetic field.}
  {\bf a}~and~{\bf b}, The energy gap $E^{\rm gap}$ 
  of the zeroth Landau levels near A ({\bf a}) or B ({\bf b}) 
  versus the strength of the magnetic field.  
  The dashed vertical line in {\bf a} indicates
  the magnetic field at which the higher- and lower-energy
  zeroth Landau levels are inverted. 
  }
  \label{Fig4}
  \end{figure}

  \begin{figure}
  \includegraphics[width=1\columnwidth]
  {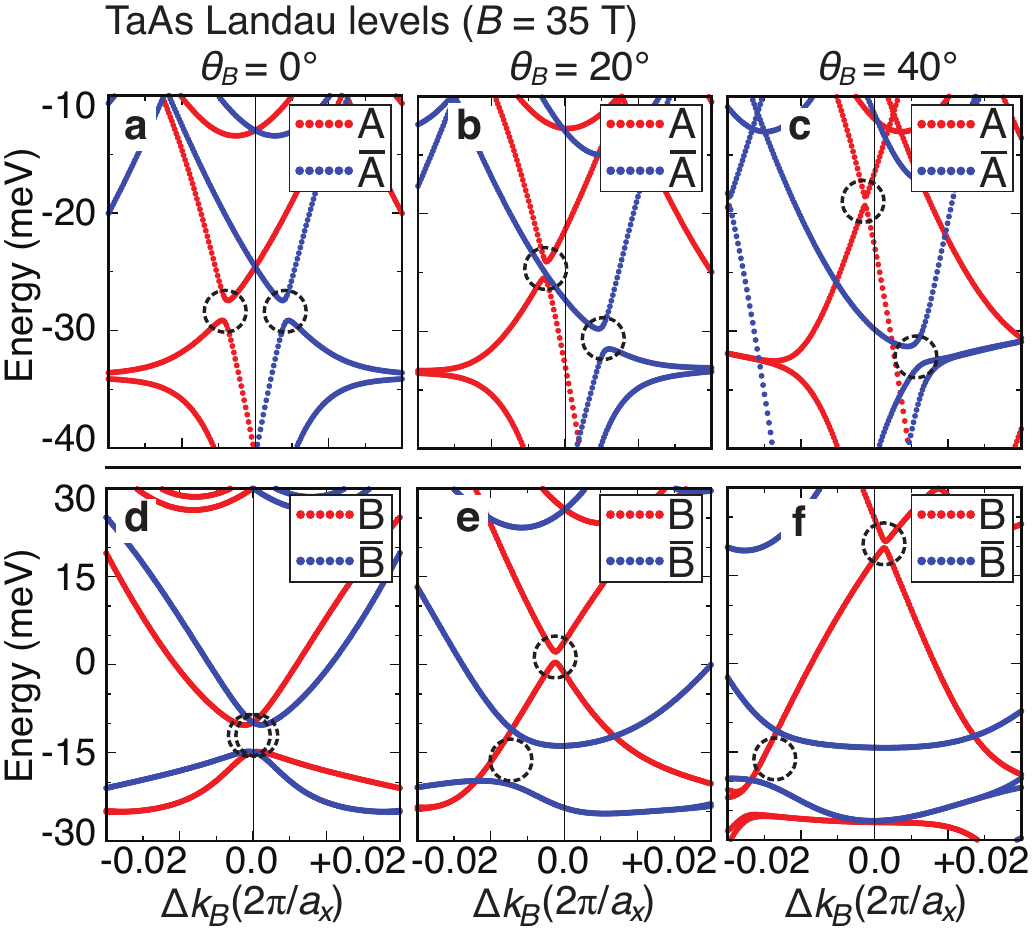}
  \caption{
  {\bf Dependence of the Landau levels of TaAs 
  on the direction of the magnetic field.}
  Landau-level spectra at $B=35$~T.  
  The $k_{x}$ component of the origin is set 
  at that of the corresponding Weyl point: 
  ${\rm A}$ or its complement $\bar{\rm A}$, {\it i.e.}, 
  ${\bf k}_{\bar{\rm A}}=-{\bf k}_{\rm A}$ ({\bf a}-{\bf c}), 
  or ${\rm B}$ or its complement $\bar{\rm B}$ ({\bf d}-{\bf f})(see Supplementary Fig. 1c~\cite{Note_supp}),  
while the $k_{y}$ component of 
the origin is set at zero.  
  The zero of energy is set at the energy of the corresponding Weyl point
  in the absence of an external magnetic field.  
  The magnetic field is applied in the $xy$ plane.
  The angle between the magnetic field and the $x$ axis, 
  $\theta_{B}$ , is $0^{\circ}$ ({\bf a} and {\bf d}), 
  $20^{\circ}$ ({\bf b} and {\bf e}), and $40^{\circ}$ ({\bf c} and {\bf f}). 
  $\Delta k_{B}$ is the distance from the origin of the wavevector 
  along the direction parallel to the applied magnetic field.
  }
  \label{Fig5}
  \end{figure}

The size of these energy gaps depends 
on the strength of the magnetic field
in a non-trivial way.
In the case that $B\le 10$~T, the energy gaps
are finite but smaller than $0.1$~meV (Fig.~4).
The energy gap increases with the magnetic field 
and could become larger than 10~meV at $B=35$~T
(Figs.~4 and~5). 
As the magnetic field further increases, 
the size of the energy gap of the ZLLs near A 
decreases, becomes zero, and then increases again (Fig.~4a).
From the calculated overlaps between the ZLL wavefunctions 
before and after the gap closure (not shown),
we find that the higher- and lower-energy ZLLs
are inverted at the gap closure.
On the other hand, the size of the energy gap 
of the ZLLs near B increases monotonically with $B$ (Fig.~4b).

For comparison, we have calculated the Landau levels 
from a linear energy-vs-wavevector model devoid of coupling 
between Weyl points with opposite chiral charges 
(see Methods~\cite{Note_supp}).
Regardless of the strength of the magnetic field,
the ZLLs obtained from the low-energy theory 
exhibit gapless, linear energy-vs-wavevector dispersions 
and well-defined chiralities (Fig.~3).
Also, the low-energy theory cannot  
properly describe the wavefunctions
of the gapped ZLLs at high magnetic fields
(Supplementary Fig.~3~\cite{Note_supp}).

We find that the size of the energy gap is 
closely related to the distance between 
the two associated Weyl points along the 
direction perpendicular to the the applied magnetic field.
(Note that the semiclassical electron orbit in momentum space
is confined in the plane perpendicular to the magnetic field.)
The two Weyl points at ${\rm A}$ %
are very close to each other (Fig.~2a) 
and the associated ZLLs can be 
easily mixed and gapped 
by a magnetic field along $x$ (Fig.~3d).
Although there are four Weyl points at ${\rm B}$ (Fig.~2a),  %
the four points are paired into two groups 
that are far away from each other 
in $k_{z}$~\cite{Huang:2015ic,PhysRevX.5.011029}, 
each consisting of two Weyl points with 
opposite chiralities that are very close together
in the Brillouin zone (Fig.~2a). 
Thus the gap opening at ${\rm B}$ (Fig.~3e) can be understood 
similarly to the case of ${\rm A}$. 
The Weyl points at ${\rm C}$ %
are far away from each other 
in the Brillouin zone (Fig.~2a); 
hence, the associated ZLLs are not mixed,
and the spectrum remains gapless (Fig.~3f).

It is tempting to conclude that a sizable
energy gap opens up if the inverse magnetic length
$l_B^{-1}=\sqrt{|e|B / \hbar }$, 
where $e$ is the charge of an electron and 
 $\hbar$ is the reduced Planck constant, 
is comparable to the the distance in momentum space
between the two Weyl points with opposite chiralities.
To the contrary,
the energy gap at a given magnetic field
is much larger at ${\rm B}$ than at ${\rm A}$ (Fig.~\ref{Fig4})
although the distance between the Weyl points at ${\rm B}$ 
is longer than twice that at ${\rm A}$.
We find that the electronic states with wavevectors
whose distance from one
Weyl point is within the distance between
the corresponding opposite-chirality Weyl points
(especially if their energies are 
close to the Weyl point energy)
play an important role in the gap opening; hence,
first-principles methods 
taking the full electronic structure into account
without assuming any specific form
of the low-energy Hamiltonian 
are useful (see Supplementary
Fig.~3 and Supplementary Discussion~1~\cite{Note_supp}).

For a comprehensive understanding of the subject matter, 
we have probed the dependence of the Landau levels 
on the direction of the magnetic field.
If the magnetic field is applied along $x$ 
the Hamiltonian is invariant under $x \rightarrow -x$.
Because of this mirror symmetry,
the Landau-level spectra at 
$k_{B}$ and $-k_{B}$ are the same (Figs.~5a and 5d).
If the magnetic field is not along $x$, those spectra are not the same 
and can be significantly different
(Figs.~5b, 5c, 5e, and 5f).

We have further studied other materials in TaAs family.
Our calculations on NbAs, TaP, and NbP (Supplementary Figs.~4-6~\cite{Note_supp})
show that the magnetic-field-induced 
chiral-symmetry breaking and gap opening generally 
occurs in Weyl semimetals.

The concept of chiral anomaly has been useful 
in understanding a variety of phenomena in Weyl semimetals. 
However, the gap opening in the zeroth Landau levels
found in our study indicates the failure of the traditional picture
of the chiral anomaly in Weyl semimetals
in a strong magnetic field.
Also, this change in the electronic structure of a Weyl semimetal
induced by a magnetic field
may lead to even more intriguing and richer physics.
Among possible mechanisms such as strain that can break
the chirality of Weyl fermions in condensed matter systems,
magnetic field is unique in that it is an inherent element
of the chiral magnetic effect initially conceived
as the signature of chiral anomaly by Nielsen and
Ninomiya~\cite{NIELSEN1983389}.

To the best of our knowledge, 
this work is the first {\it ab initio} study on  
the Landau levels of real Weyl semimetals
that takes account of the full electronic structure.
The key findings of our study
are beyond the description of
the commonly adopted low-energy theories, 
in which the coupling between Weyl points 
by a magnetic field is assumed to be 
negligibly weak without justification.
The powerful method used in our study
can be applied in investigating the electromagnetic responses in
other classes of
novel semimetallic phases.

{\it Note added}:
After we finished this work, 
we became aware of two independent experimental studies reporting  
that the longitudinal magnetoresistance
of TaAs at a high-field regime increases with the strength of
the magnetic field and becomes positive~\cite{zhang2017magneticm,2017Ramshawm}.
In Ref.~\cite{zhang2017magneticm}, although the authors did not discuss the mechanism in detail
in their preprint which does not have the supplementary
information, they ascribed the experimental results
to the gapping of the Weyl points at strong magnetic fields.
After submission of our manuscript, a paper reporting
an experimental discovery of
a sharp sign reversal in the Hall resistivity of TaP at
$B=35$~T was published~\cite{Zhang:2017jg}.
The authors attributed the results to an energy gap opening
at the Weyl points due to the magnetic tunneling at strong
magnetic fields.
They have introduced a $\mathbf{k}\cdot\mathbf{p}$
Hamiltonian which can describe the region in k space
close to the Weyl
points to explain their experimental data.
Lastly, after this work was made public through posting as an e-print~\cite{Kim:2017arm},
two studies relevant to the breakdown of the chiral anomaly
also became public as e-prints~\cite{PhysRevB.96.195143,Saykin:2017arm}.
In Ref.~\cite{PhysRevB.96.195143}, Chan and Lee
reported that the anisotropy in group velocity
near the Weyl point plays an important role in the hybridization
and gapping between the zeroth Landau levels
by using an effective Hamiltonian.
Our results that the energy gap in Landau level spectrum is much
larger at {\rm B} than at {\rm A} (Fig.~\ref{Fig4})
are consistent with their theory when applied to TaAs.

\begin{acknowledgments}
This work was supported by the Creative-Pioneering Research Program through Seoul
National University.
\end{acknowledgments}

\bibliography{manuscript}

\newpage

\beginsupplement

\begin{widetext}
% \section*{Supplemental Material: Breakdown of the Chiral Anomaly in Weyl Semimetals in a Strong Magnetic Field} 
{\Large {\bf Supplemental Material: Breakdown of the Chiral Anomaly in Weyl Semimetals in a Strong Magnetic Field} }
\end{widetext}

\subsection{Methods}

 {\bf Electronic-structure calculations.}  
 Electronic structures were calculated within 
 the framework of density functional theory 
 as implemented 
 in the {\tt Quantum-ESPRESSO} package~\cite{QE_cite}.  
 Spin-orbit coupling effects were treated self-consistently 
 using fully-relativistic, norm-conserving 
 pseudopotentials~\cite{PhysRevB.88.085117,SCHLIPF201536}.  
 The exchange-correlation energy 
 was approximated by the scheme of 
 Perdew, Burke, and Ernzerhof~\cite{PhysRevLett.77.3865}.  
 The kinetic energy cutoff was set to 100 Ry, 
 and the Brillouin zone was sampled 
 with a $ 12 \times 12 \times 8 $ Monkhost-Pack~\cite{PhysRevB.13.5188} 
 k-point mesh.  
 The experimental lattice parameters~\cite{Boller:a04000,WILLERSTROM1984273,
 Huang:2015ic,Xue1501092}  were used.

\bigbreak

 {\bf{\it Ab initio} tight-binding model.  }
 We constructed maximally-localized 
 Wannier functions using the {\it d} orbitals 
 of Ta or Nb atoms and {\it p} orbitals of As or P atoms 
 as an initial guess~\cite{PhysRevB.56.12847,PhysRevB.65.035109,MOSTOFI2008685,RevModPhys.84.1419}.  
 We then obtained the Hamiltonian matrix elements 
 between the Wannier orbitals.  
 We incorporated the effect of a magnetic field 
 into the tight-binding Hamiltonian 
 via the Peierls substitution~\cite{Peierls1933}, 
 \begin{equation} 
 t_{ij}\rightarrow t_{ij} \exp \left[ i \frac{e}{\hbar} 
 \int_{{\mathbf r}_{i}}^{{\mathbf r}_{j}}
 {\mathbf A}\cdot d{\mathbf l} \right],
 \label{Peierls}
 \end{equation}
 where $e$ is the charge of an electron, 
 $\hbar$ is the reduced Planck constant, 
 $t_{ij}$ is the Hamiltonian matrix element 
 between the Wannier orbitals $i$ and $j$
 in the absence of an external magnetic field, 
 and ${\mathbf A}=\left( B_{y}z,-B_{x}z,0 \right)$ 
 is the vector potential that 
 corresponds to the magnetic field 
 ${\mathbf B}=\left( B_{x},B_{y},0 \right)$.
 The integration was performed along 
 a straight line between the Wannier centers 
 ${\mathbf r}_{i}$ and ${\mathbf r}_{j}$ 
 following the theory of Ref.~\cite{PhysRevLett.87.087402}.  
 We have checked that our computational results 
 do not depend on the gauge of the vector potential.  
 As the additional phase in Eq.~\eqref{Peierls} breaks 
 the translational symmetry, 
 we employed the supercell method. 
 Nielsen and Ninomiya~\cite{NIELSEN1983389} and the recent studies 
 on the Landau levels of Weyl semimetals~\cite{PhysRevLett.117.077202,PhysRevLett.117.086401,PhysRevLett.117.086402} 
 have neglected the effects of Zeeman splitting 
 because (i) the orbital effects 
 are more important~\cite{PhysRevLett.117.077202,PhysRevLett.117.086401,PhysRevLett.117.086402} and, 
 (ii) much more importantly, 
 the Zeeman splitting merely 
 induces a shift in the positions of the Weyl points 
 in the Brillouin zone
 and hence results only in quantitative changes 
 and all the qualitative findings 
 in those studies~\cite{NIELSEN1983389,PhysRevLett.117.077202,PhysRevLett.117.086401,PhysRevLett.117.086402} 
 and in ours remain valid.
 
 \bigbreak
 
 {\bf{\it Ab initio} $2 \times 2$ model. }  
 Near a Weyl point, 
 the effective, low-energy Hamiltonian 
 at wavevector ${\mathbf k}$ can be written as 
 $H \left({\mathbf k} \right) = \sum_{i=1}^{3}\hbar k_{i} V_{i} I + \sum_{i,j=1}^{3} \hbar k_{i} A_{ij} \sigma_{j},$
 where $I$ is the two-dimensional identity matrix 
 and $\sigma$\textrm{'}s are the Pauli matrices.  
 Previous studies have determined 
 the parameters in the effective Hamiltonian 
 to best reproduce the {\it ab initio} 
 electronic band structure near 
 the Weyl point~\cite{Soluyanov:2015cn,PhysRevLett.117.066402,PhysRevB.93.205133,PhysRevB.94.121117,1367-2630-19-3-035001}.  
 In our study, we took advantage of 
 the matrix elements of the velocity operator 
 right at the Weyl point 
 between the doubly-degenerate 
 eigenstates~\cite{PhysRevB.74.195118,PhysRevB.75.195121}.  
 The velocity matrix elements 
 are written as $v_{nm,i}=\braket{\psi_{n}|\hat{v}_i|\psi_{m}}$, 
 where $\hat{v}_i$ is the $i$-th component of the velocity operator,
 $m$ and $n$ are the band indices, and
 $\ket{\psi_{1}}$ and 
 $\ket{\psi_{2}}$ are 
 the degenerate Bloch states 
 at the Weyl point.  
 In terms of the velocity matrix elements, 
 $V_{i}$ and $A_{ij}$ are given by 
 \begin{equation} 
 \begin{split}
 V_{i} &= \left( v_{11,i}+v_{22,i} \right) / 2 \\ 
 A_{i1} &= {\rm Re}\left[ v_{12,i} \right] \\
 A_{i2} &= -{\rm Im}\left[ v_{12,i} \right] \\
 A_{i3} &= \left( v_{11,i}-v_{22,i} \right)/2.
 \end{split}
 \end{equation}
 Unlike the previous methods based on 
 {\it ab initio} electronic band structures~\cite{Soluyanov:2015cn,PhysRevLett.117.066402,PhysRevB.93.205133,PhysRevB.94.121117,1367-2630-19-3-035001}, 
 our method naturally determines 
 the sign of the chiral charge 
 of a Weyl point, which is the sign of 
 the determinant of matrix $A$.  
 The effect of the magnetic field 
 was incorporated into the Hamiltonian  
 $H \left( {\mathbf k} \right)$
 by the minimal coupling,  
 $H \left( {\mathbf k} \right) \rightarrow 
 H \left( {\mathbf k} -\frac{e}{\hbar} {\mathbf A} \right)$.  
 For numerical calculations, 
 we employed the supercell method 
 as in the {\it ab initio} tight-binding model.

\newpage

  \begin{figure*}
  \begin{center}
  \includegraphics[width=1.4\columnwidth]
  {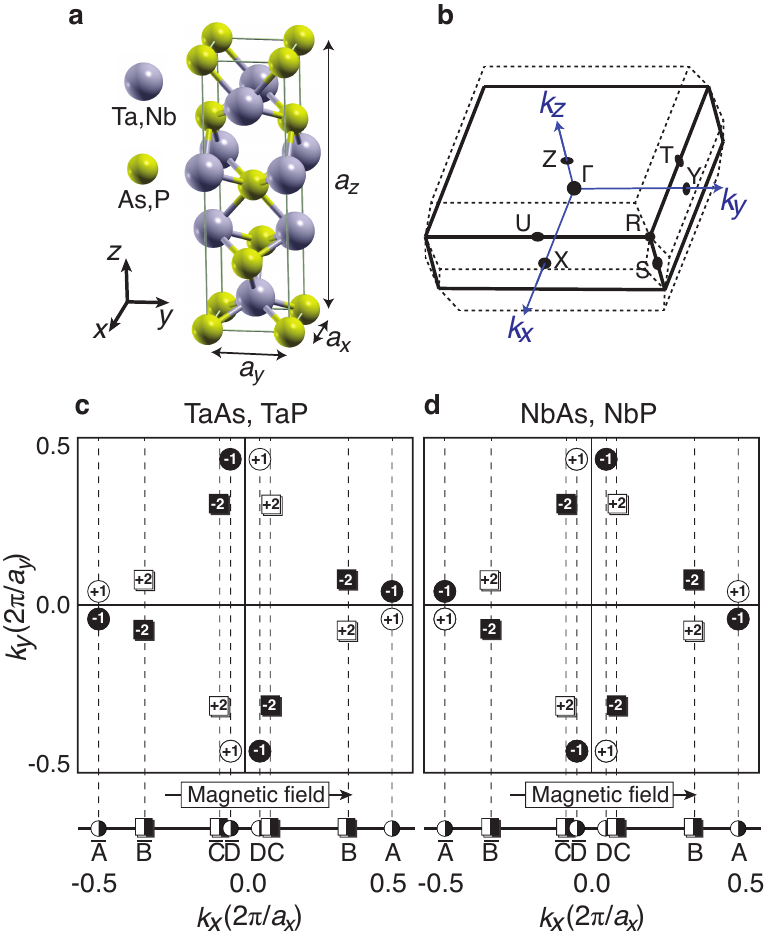}
  \end{center}
  \caption{{\bf Crystal structure and Brillouin zone 
  of TaAs, NbAs, TaP, and NbP.}
  {\bf a}, Crystal structure of TaAs, NbAs, TaP, and NbP. 
  $a_{x}$, $a_{y}$, and $a_{z}$ are the lattice constants
  of the tetragonal (conventional) unit cell 
  along the $x$, $y$, and $z$ directions ($a_{x}=a_{y}$), respectively.
  {\bf b}, Brillouin zone of TaAs, NbAs, TaP, and NbP.  
  The Brillouin zone of the tetragonal (conventional) unit cell, 
  which is used in the computations, 
  is drawn in thick solid lines.  
  The Brillouin zone of the body-centered tetragonal (primitive) 
  unit cell is drawn in thin dashed lines.  
  {\bf c} and {\bf d}, A schematic (not to scale) 
  of the projected Weyl points 
  and their chiral charges (enclosed numbers) 
  of TaAs or TaP ({\bf c}) and NbAs or NbP ({\bf d}).  
  }
  \label{SuppFig1}
  \end{figure*}

  \begin{figure*}
  \begin{center}
  \includegraphics[width=1.4\columnwidth]
  {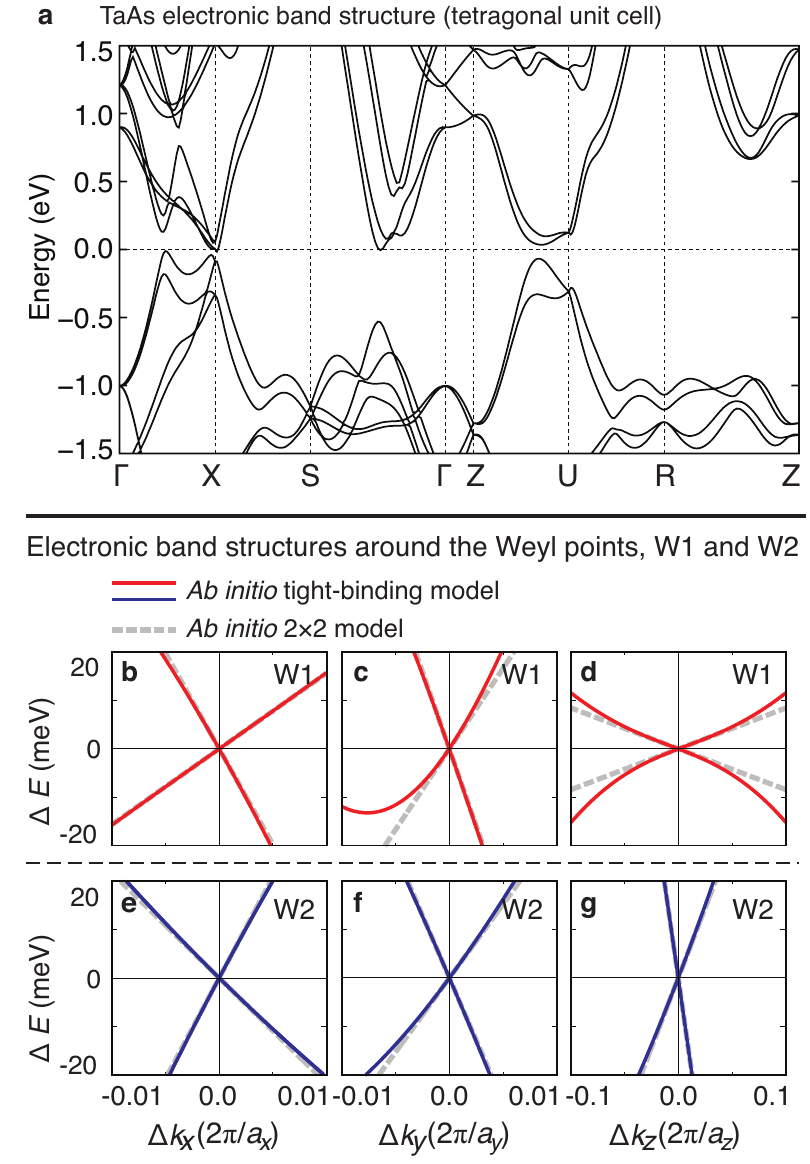}
  \end{center}
  \caption{{\bf Electronic band structures of TaAs.}
  {\bf a}, Electronic band structure along 
  the symmetry lines of TaAs shown in Supplementary Fig.~1.  
  The zero of energy is at the Fermi level.  
  {\bf b}-{\bf g}, 
  Electronic band structures along $k_{x}$ ({\bf b} and {\bf e}), 
  $k_{y}$ ({\bf c} and {\bf f}), and $k_{z}$ ({\bf d} and {\bf g}).  
  The origins of wavevector are set at the Weyl points,
  W$_{1}~(0.4879 \cdot 2\pi/a_{x}$, $0.0077 \cdot 2\pi/a_{y}$, $0$) 
  ({\bf b}-{\bf d}) 
  and W$_{2}~(0.2812 \cdot 2\pi/a_{x}$, $0.0197 \cdot 2\pi/a_{y}$, 
  $0.4067 \cdot 2\pi/a_{z}$) ({\bf e}-{\bf g}). 
  The zeros of energy are set at 
  the energies of the corresponding Weyl points.  
  Solid red or solid blue curves represent the computational results 
  obtained from the {\it ab initio} tight-binding model.  
  Dashed grey curves show the results from the {\it ab initio} $2\times2$ model.  
  }
  \label{SuppFig2}
  \end{figure*}

  \begin{figure*}
  \begin{center}
  \includegraphics[width=1.6\columnwidth]
  {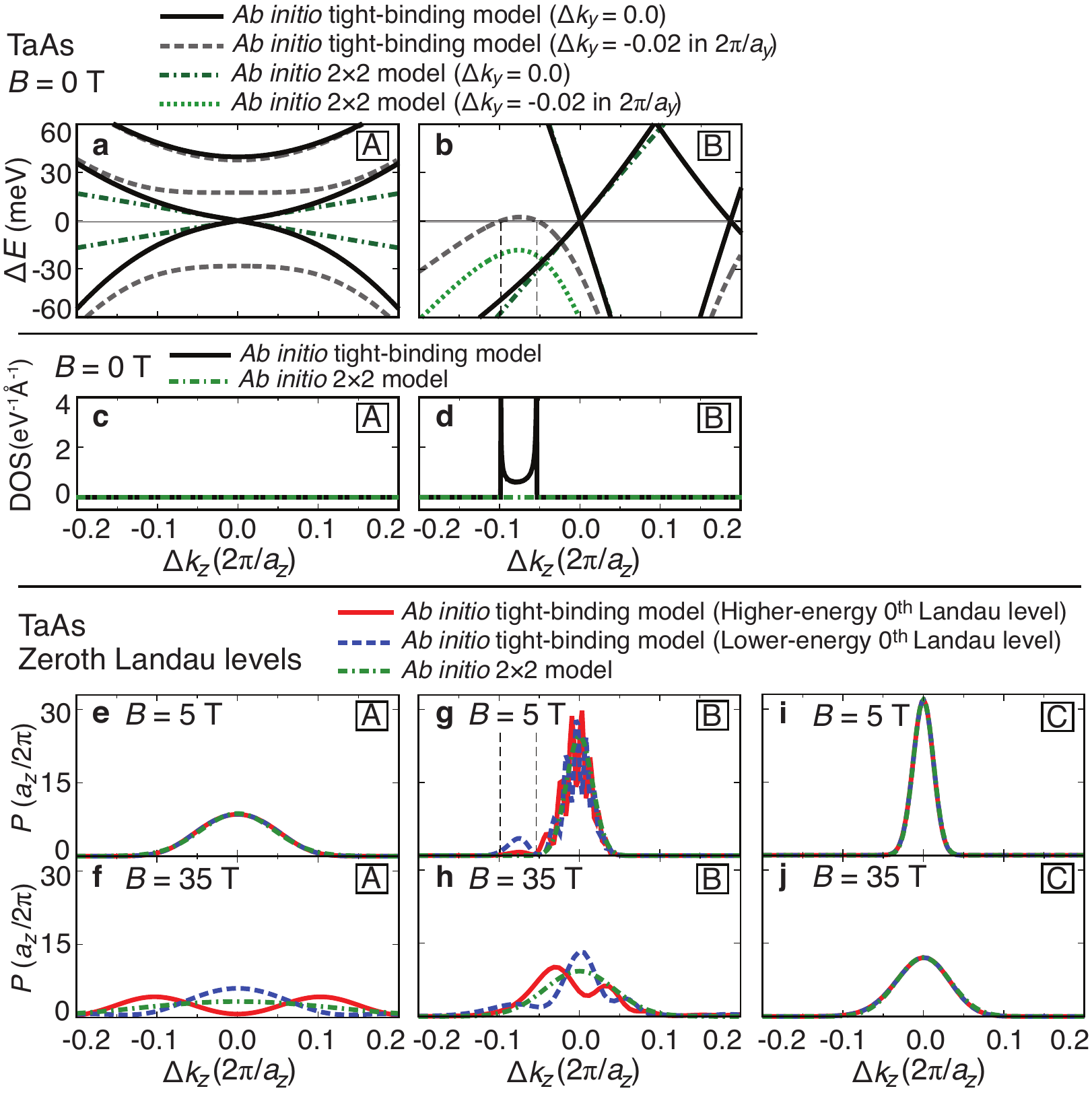}
  \end{center}
  \caption{{\bf Zeroth-Landau-level states of TaAs.}
  {\bf a} and {\bf b}, Electronic band structures along ${\it k_{z}}$ 
  for two different ${\it k_{y}}$ values.  
  The origins of wavevector are set at the Weyl points, W$_{1}$ ({\bf a}) 
  and W$_{2}$ ({\bf b}) (see the caption of Supplementary Fig.~2).  
  The zeros of energy are set at the energies of the corresponding Weyl points.
  Solid black or dashed grey curves show the results from the {\it ab initio}
  tight-binding model. 
  Dash-dotted dark green and dotted green curves show the results from the 
  {\it ab initio} $2\times2$ model.
  {\bf c} and {\bf d}, 
  The $k_{z}$-resolved density of states
  whose energy and $k_{x}$ 
  are the same as those of the Weyl points
  at W$_{1}$ ({\bf c}) and W$_{2}$ ({\bf d}):
%   \begin{equation} 
$
 {\rm DOS}\left(k_{z}\right)=\frac{1}{2\pi}
 \sum_{n}^{} \int_{\rm BZ} d\mathbf{k}'
 \delta \left( E_{n,\mathbf{k}'} - E^{\rm W} \right)
\delta \left( k'_{x}-k_{x}^{\rm W} \right)
 \delta \left( k'_{z}-k_{z} \right)
$
%  \end{equation}
  where the $E_{n,\mathbf{k}'}$ is the energy of the 
  Bloch states at wavevector $\mathbf{k}'$ with band index $n$,
  $E^{\rm W}$ and $k_{x}^{\rm W}$ are 
  the energy and $k_{x}$ of the Weyl point, respectively,
  the integration is carried out over the Brillouin zone, 
  and the summation runs over all band indices $n$.
  The dashed black lines in {\bf b} and {\bf g} are guides to the eye
  indicating the van Hove singularities.
  {\bf e}-{\bf j}, The projection $P$ of the zeroth-Landau-level states 
  at ${\rm A}$ ({\bf e} and {\bf f}), ${\rm B}$ ({\bf g} and {\bf h}), 
  and ${\rm C}$ ({\bf i} and {\bf j}) 
  at $B=5$~T ({\bf e}, {\bf g}, and {\bf i}) 
  and $B=35$~T ({\bf f}, {\bf h}, and {\bf j}) 
  onto the Bloch states in 
  the absence of a magnetic field. 
  The projection $P$ is given by 
%  \begin{equation} 
$
 P_{\pm}\left(k_{z}\right)=\frac{2\pi}{a_{z}}
 \sum_{n}^{} 
 \lvert \braket{\psi_{n,k_{x}^{\rm W}}\left(k_{z}\right)
 |\psi_{\pm,k_{x}^{\rm W}}^{\rm L}} \rvert ^{2} 
$
%  \end{equation}
  where 
  $\ket{\psi_{+}^{\rm L}}$ 
  and $\ket{\psi_{-}^{\rm L}}$ are the higher- and 
  lower-energy zeroth-Landau-level states, respectively,
  and $\ket{\psi}$ is the Bloch state in 
  the absence of a magnetic field. 
  The summation runs over all band indices of the Bloch states $n$,
  and $k_{x}^{\rm W}$ is the $k_{x}$ component 
  of the corresponding Weyl point.
  Solid red and dashed blue curves 
  show $P_{+}\left(k_{z}\right)$ and $P_{-}\left(k_{z}\right)$, respectively, 
  obtained from the {\it ab initio} tight-binding model.  
  Dash-dotted green curves show  the average of 
  $P_{+}\left(k_{z}\right)$ and $P_{-}\left(k_{z}\right)$
  obtained from the {\it ab initio} $2\times2$ model;
  we took their average because the two zeroth-Landau-level states 
  are degenerate.
  The origins of $k_{z}$ are 
  the $k_{z}$ components of W$_{1}$({\bf e} 
  and {\bf f}) and W$_{2}$({\bf g}-{\bf j}).
  }
  \label{SuppFig3}
  \end{figure*}

\newpage

\subsection{Supplementary Discussion 1: Discussion on
the results shown in Supplementary Fig. 3}

   There are two pronounced peaks in the density of states
  near the Weyl point ${\rm B}$ only in the full {\it ab-initio}
  tight-binding 
  calculations but not in {\it ab initio} $2\times2$ model calculations (Supplementary Fig.~3d).
  From perturbation theory,
  nearby Bloch states with similar energies as $E^{\rm W}$
  strongly mix by the magnetic field and form Landau-level states.  
  We emphasize that in order to describe this mixing correctly,
  {\it ab initio} calculations taking into account the full electronic
  states should be used 
  because the states whose Bloch wavevectors are within the range of the separation
  between the two opposite-chirality Weyl points are all mixed together by an external magnetic field
  and are of crucial importance in forming the Landau levels and in opening up the band gap at the
  zeroth Landau levels.
  In our case, the distance of the van Hove singularities from
  the Weyl point at B ($\sim0.065$~\AA$^{-1}$) (Supplementary Fig.~3b) is
  less than the separation of the two opposite-chirality Weyl points
  ($=0.072$~\AA$^{-1}$). 
  Therefore the van Hove singularity
  at energy $E^{\rm W}$ (Supplementary Figs.~3b and 3d)
  plays an important role in the gap opening.
    
   The projection $P_-(k_z)$ from the {\it ab initio} tight-binding
  model at ${\rm B}$ is sizable
  near the van Hove singularities (Supplementary Fig.~3g).
        As we discussed above, the van-Hove-singularity states
  indeed participate in forming the zeroth-Landau-level states.
  Note that the {\it ab initio} $2\times2$ model cannot properly describe 
  the wavefunctions of the zeroth Landau levels if gapped
  (Supplementary Figs.~3f-3h; also see Fig.~3 of the main manuscript).

  We conclude by emphasizing the importance of handling the
  full electronic states whose wavevector is within the range of
  the separation between the two opposite-chirality Weyl points 
  in describing this gap-opening physics.
  Our proposed {\it ab initio} tight-binding method is
  well suited for this purpose.

  \begin{figure*}
  \begin{center}
  \includegraphics[width=2.0\columnwidth]
  {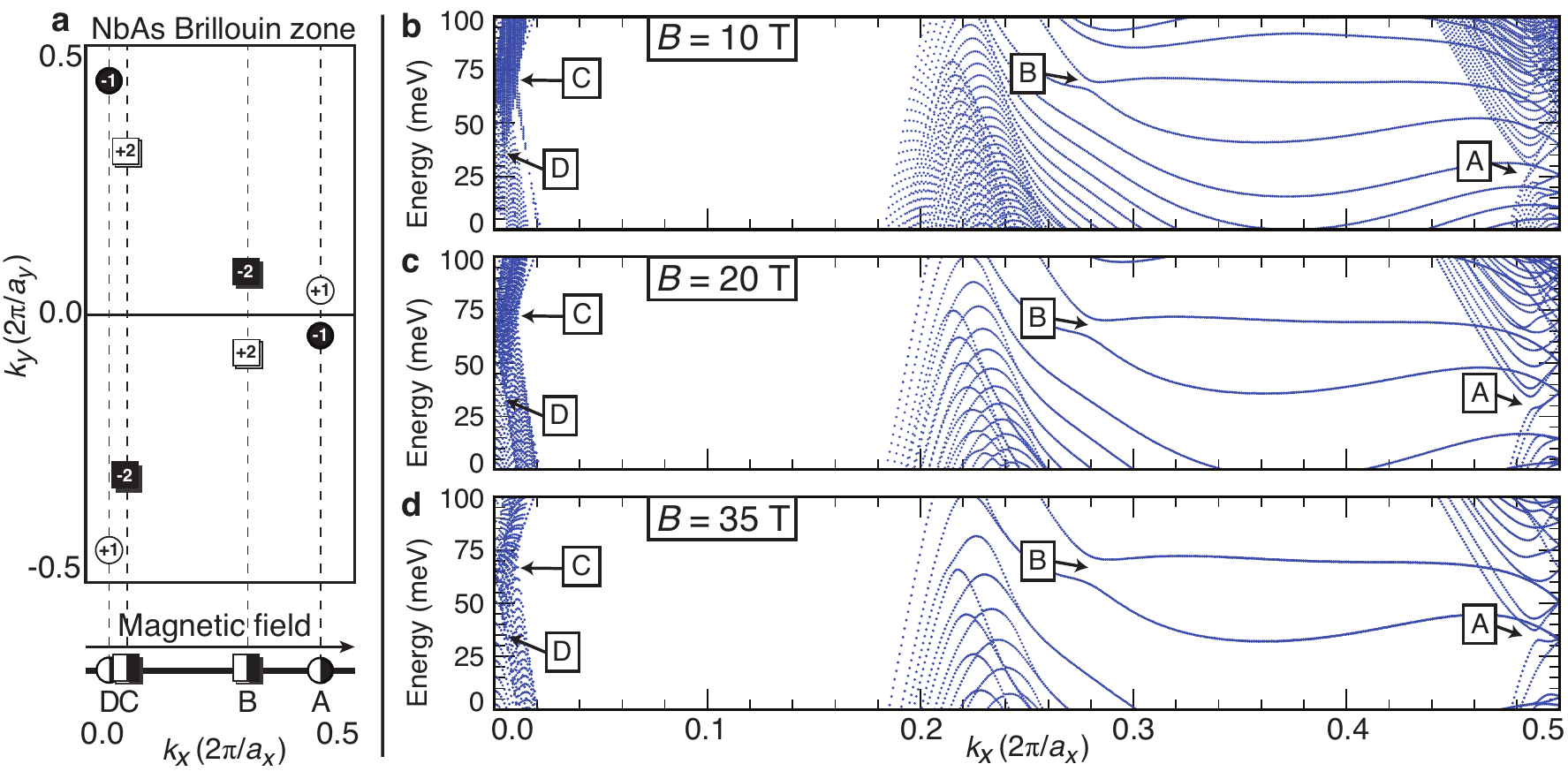}
  \end{center}
  \caption{{\bf Landau levels of NbAs.}
  {\bf a}, A schematic (not to scale) of the projected Weyl points 
  and their chiral charges (the enclosed numbers).  
  Half of the Brillouin zone ($k_{x}>0$) is shown.  
  Discs and squares represent two different classes of Weyl points.  
  {\bf b}-{\bf d}, Landau-level spectra of NbAs 
  in a magnetic field applied along $x$ 
  obtained from {\it ab~initio} tight-binding calculations.  
  The zero of energy is set at the Fermi level before the application of the magnetic field.
  ${\rm A}$, ${\rm B}$, ${\rm C}$, and ${\rm D}$ denote the corresponding Weyl points shown in {\bf a}.  
  }
  \label{SuppFig4}
  \end{figure*}

  \begin{figure*}
  \begin{center}
  \includegraphics[width=2.0\columnwidth]
  {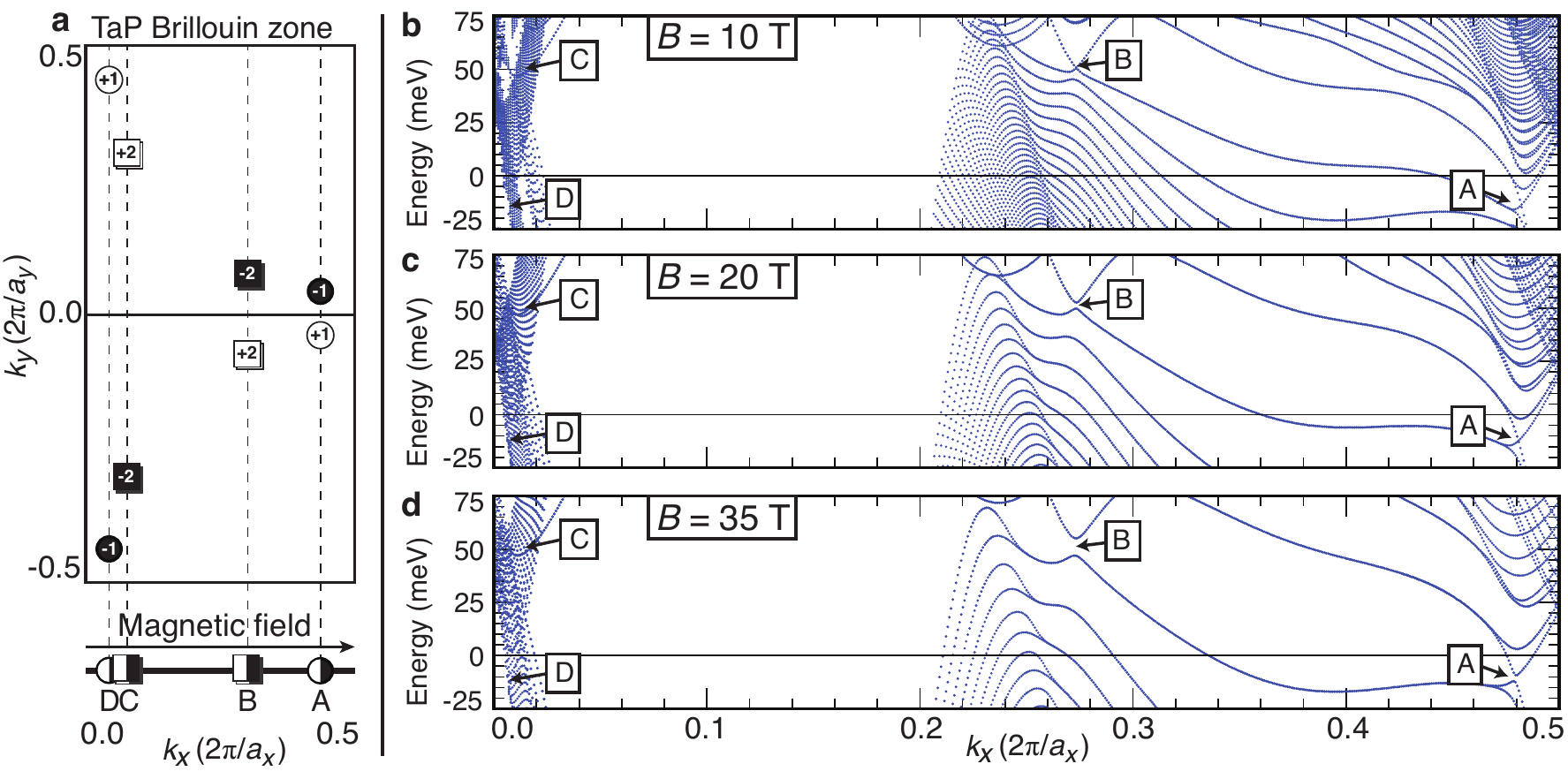}
  \end{center}
  \caption{{\bf Landau levels of TaP.}
  {\bf a}, A schematic (not to scale) of the projected Weyl points 
  and their chiral charges (the enclosed numbers).  
  Half of the Brillouin zone ($k_{x}>0$) is shown.  
  Discs and squares represent two different classes of Weyl points.  
  {\bf b}-{\bf d}, Landau-level spectra of TaP 
  in a magnetic field applied along $x$ 
  obtained from {\it ab~initio} tight-binding calculations.  
  The zero of energy is set at the Fermi level before the application of the magnetic field.
  ${\rm A}$, ${\rm B}$, ${\rm C}$, and ${\rm D}$ denote the corresponding Weyl points shown in {\bf a}.  
  }
  \label{SuppFig5}
  \end{figure*}

  \begin{figure*}
  \begin{center}
  \includegraphics[width=2.0\columnwidth]
  {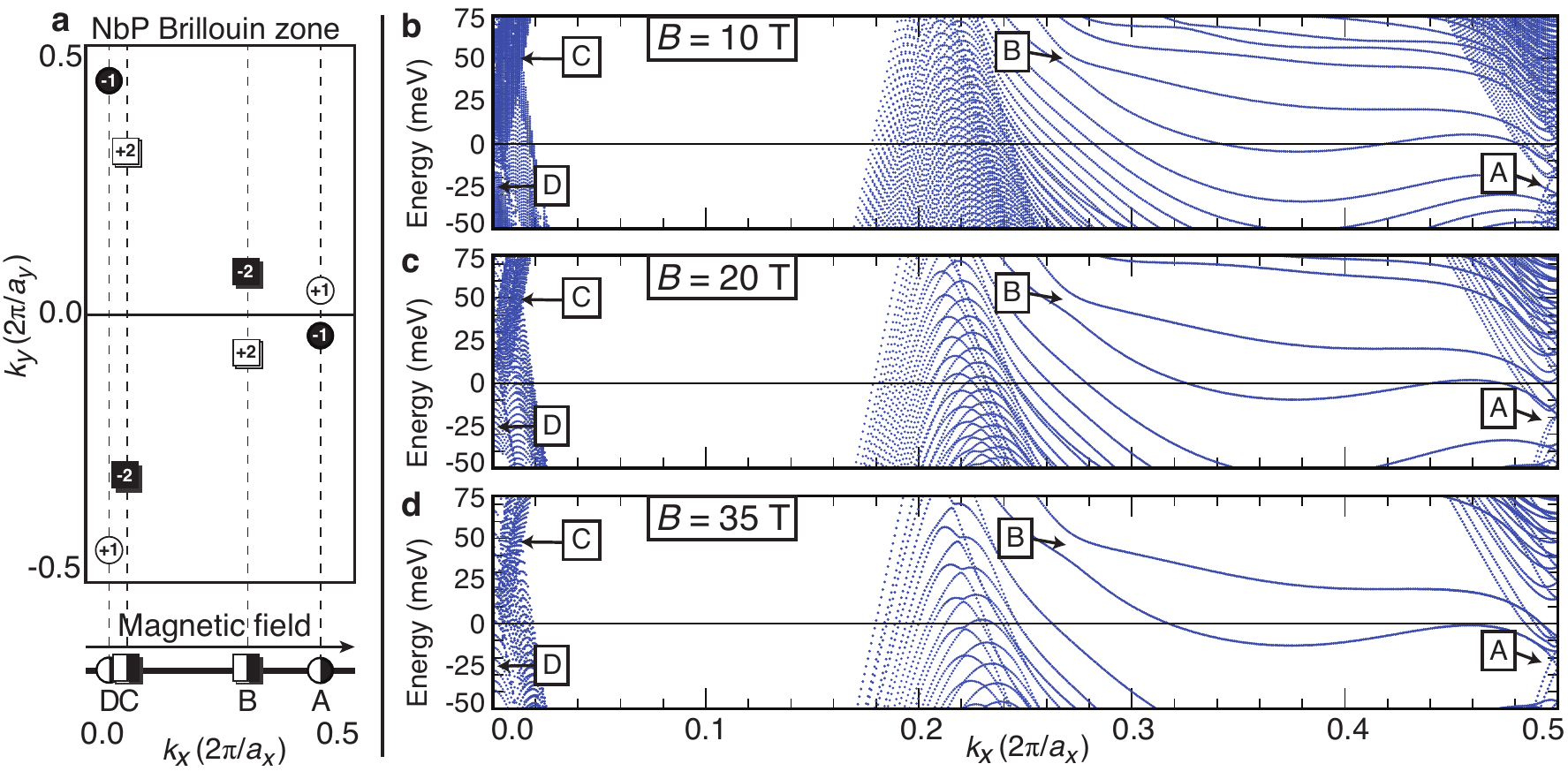}
  \end{center}
  \caption{{\bf  Landau levels of NbP.}
  {\bf a}, A schematic (not to scale) of the projected Weyl points 
  and their chiral charges (the enclosed numbers).  
  Half of the Brillouin zone ($k_{x}>0$) is shown.  
  Discs and squares represent two different classes of Weyl points.  
  {\bf b}-{\bf d}, Landau-level spectra of NbP 
  in a magnetic field applied along $x$ 
  obtained from {\it ab~initio} tight-binding calculations.  
  The zero of energy is set at the Fermi level before the application of the magnetic field.
  ${\rm A}$, ${\rm B}$, ${\rm C}$, and ${\rm D}$ denote the corresponding Weyl points shown in {\bf a}.  
  }
  \label{SuppFig6}
  \end{figure*}

\end{document}